\documentclass[aps,prl,twocolumn,showpacs,preprintnumbers,amsmath,amssymb, superscriptaddress]{revtex4}

\usepackage{graphicx}
\usepackage{dcolumn}
\usepackage{bm}

\begin{document}

\preprint{APS/123-QED}

\title
{
Beating the standard quantum limit: \\
Phase super-sensitivity of N-photon interferometers
}

\author{Ryo Okamoto}
\altaffiliation[Electronic address: ]{oka@es.hokudai.ac.jp}
\affiliation{%
Research Institute for Electronic Science, Hokkaido University,
Sapporo 060--0812, Japan
}%

\author{Holger F. Hofmann}
\affiliation{%
Graduate School of Advanced Sciences of Matter, Hiroshima University Hiroshima 739-8530, Japan
}%

\author{Tomohisa Nagata}
\affiliation{%
Research Institute for Electronic Science, Hokkaido University,
Sapporo 060--0812, Japan
}%

\author{Jeremy L. O'Brien}
\affiliation{Centre for Quantum Photonics, H. H. Wills Physics Laboratory \& Department of Electrical and Electronic Engineering, University of Bristol, Merchant Venturers Building, Woodland Road, Bristol, BS8 1UB, UK
}%

\author{Keiji Sasaki}
\affiliation{%
Research Institute for Electronic Science, Hokkaido University,
Sapporo 060--0812, Japan
}%

\author{Shigeki Takeuchi}
\affiliation{%
Research Institute for Electronic Science, Hokkaido University,
Sapporo 060--0812, Japan
}%

\date{\today}
%

\begin{abstract}
Quantum metrology promises greater sensitivity for optical phase measurements than could ever be achieved classically. Here we present a theory of the phase sensitivity for the general case where the detection probability is given by an $N$ photon interference fringe. We find that the phase sensitivity has a complex dependence on both the intrinsic efficiency of detection $\eta$ and the interference fringe visibility $V$. Most importantly, the phase that gives maximum phase sensitivity is in general not the same as the phase at which the slope of the interference fringe is a maximum, as has previously been assumed. We determine the parameter range where quantum enhanced sensitivity can be achieved. In order to illustrate these theoretical results, we perform a four photon experiment with $\eta=3/4$ and $V=82\pm6$\% (an extension of our previous work [Science \textbf{316}, 726 (2007)]) and find a phase sensitivity 1.3 times greater than the standard quantum limit at a phase different to that which gives maximum slope of the interference fringe. 

\end{abstract}

\maketitle
%
\textit{Introduction}.---The subwavelength sensitivity offered by optical phase measurements is the reason that they have found applications across all fields of science, from cosmology (gravitational detection) to nanotechnology (phase-contrast microscopy). Given finite resources (energy, number of photons, etc) the phase sensitivity is limited by statistical uncertainty. It has been shown that the use of semi-classical probes, \textit{i.e.} coherent light fields, limits the sensitivity to the standard quantum limit (SQL): $\Delta\phi=1/\sqrt{N}$, where $N$ is the average number of photons used when the probe interacts only once with the phase-changing object \cite{vit-prl-96-010401}. The more fundamental Heisenberg limit is attainable with the use of a quantum probe (eg. an entangled state of photons): $\Delta\phi=1/N$ \cite{ca-prd-23-1693, yu-pra-33-4033, vit-sci-306-1330, vit-prl-96-010401}. The possibility to beat the SQL and approach the Heisenberg limit is therefore of great fundamental interest in understanding how quantum effects can be advantageous and may lead to important applications in the precision measurements that are the basis of all quantitative science.

In this context, interference experiments using two- \cite{ou-pra-42-2957,ra-prl-65-1348,ku-qso-10-493,fo-prl-82-2868,ed-prl-89-213601,ei-prl-94-090502}, three- \cite{mi-nat-429-161}, and four-photon states \cite{wa-nat-429-158, su-pra-74-033812,na-sci-316-726} have been reported. In each case $p \propto \sin(N \phi)$, where $p$ is the detection probability that gives rise to the $N$ photon interference. Observation of such a ``$\lambda/N$" fringe, with a period $N$ times shorter than the single photon fringe of a semi-classical resource, is called \textit{phase super-resolution} \cite{mi-nat-429-161,re-prl-98-223601}. Phase super-resolution has sometimes been associated with beating the SQL: \textit{phase super-sensitivity}. However, Resch \textit{et al.} have recently shown that phase super-resolution can be achieved with purely classical resources \cite{re-prl-98-223601}. In contrast, phase super-sensitivity is necessary to gain a quantum advantage in precision, and requires quantum resources.  

In their groundbreaking work, Resch \textit{et al.} pointed out that the phase sensitivity depends on both the $\lambda/N$ fringe visibility $V$ and the efficiency $\eta$ \footnote{Note that Resch \textit{et al.} used $\eta$ as a phenomenological parameter that could include various kinds of experimental efficiencies, whereas we consider just the intrinsic efficiency of the experimental scheme.}. Since a lower efficiency means that more photons are necessary to achieve a given measurement precision, they assumed that it is possible to treat the efficiency simply as an increase in the required photon number by a factor of $1/\eta$. However, this assumption requires that the statistical errors of the successful measurements do not depend on the efficiency. As we show in the following, this is not the case in typical experiments, where a particular event is selectively detected by coincident photon counts to observe a $\lambda/N$ interference fringe, since the detection probability $p$ inseparably combines the effects of $\eta$ and $V$. Therefore Resch \textit{et al.}'s theory does not apply to previous experiments with $N>2$ such as the ones reported in \cite{mi-nat-429-161, wa-nat-429-158, su-pra-74-033812, na-sci-316-726}, and a more detailed theoretical analysis is required to determine the effects of efficiency $\eta$ on the phase sensitivity of $N$ photon interference experiments.

Here we present a complete theory of the phase sensitivity $\mathcal{S}$ for the general case where the detection probability is given by an $N$ photon interference fringe. We perform a statistical error analysis for the event-detection probability which is valid for all $N$-photon interference experiments, including those reported previously \cite{mi-nat-429-161, wa-nat-429-158, su-pra-74-033812, na-sci-316-726}. We find that the phase sensitivity has a complex dependence on both the intrinsic efficiency of detection $\eta$ and the interference fringe visibility $V$. Furthermore, the phase at which $\mathcal{S}$ is a maximum is generally \textit{not} the same as the phase at which the slope of $p$ is a maximum, as has widely been assumed. In order to illustrate these theoretical results, we apply this expression for $\mathcal{S}$ to a new experiment with $N=4$, $\eta=3/4$ (improved by a factor of two from our previous experiment \cite{na-sci-316-726}) and $V=82\pm6$\%, and find  a phase super-sensitivity of 1.3 times greater than the SQL. As anticipated from our theoretical analysis, we find that the maximum phase sensitivity does not occur at the maximum slope of $p$.

%
\textit{Derivation of $\mathcal{S}$}.---To derive the phase sensitivity in a typical $N$-photon interference experiment, we start with the probability for the successful detection of the desired $N$-photon event in a single trial. For a $\lambda/N$ fringe, the phase dependence of this probability is given as
\begin{equation}
p(\phi)= \frac{\eta}{2}\left[1+V \sin(N\Phi_0 + N\phi)\right], \label{Eq:prob}
\end{equation}
where $\Phi_0$ is a bias phase and $\phi$ is the small phase shift to be measured. The interference fringe is thus characterized by two key parameters: the intrinsic efficiency $0 \le \eta \le 1$ is determined by the experimental scheme used and indicates the probability with which a given input photon contributes to the $N$-photon interference; and
the visibility $0 \le V \le 1$ of the interference fringe observed in the output of the interferometer indicates the quality of the $N$-photon interference.

Given the dependence of detection probability on phase (eq. 1), the phase sensitivity can be derived using standard methods of metrology \cite{NC-Text}. The phase estimate is determined from the average number of times that the selected $N$-photon event is observed in $k$ trials,
\begin{equation}
C_{k} = k~p(\phi).
\end{equation}
Small phase shifts can be estimated by the deviation of
$C_{k}$ from its value at $\Phi_0$. However, the measurement result has a statistical variance of
\begin{equation}
\Delta C_{k}^{2} = k~p (1-p),
\label{Eq:deltaCk}
\end{equation}
so phase shifts cannot be distinguished from statistical errors if the change of $C_{k}$ is smaller than $\Delta C_{k}$. Specifically, the error of the phase estimate is given by
\begin{equation}
\delta \phi^2
= \frac{\Delta C_{k}^{2}}{(dC_{k}/d\phi)^2}.
\label{EQ:phUn}
\end{equation}
For direct comparison with the SQL, we can define the phase sensitivity as the square root of the ratio of this phase error and the phase error of $1/\sqrt{N}$ at the SQL:
\begin{eqnarray}
\mathcal{S}^2 &\equiv& \left( kN \delta \phi^2 \right) ^{-1}
\nonumber \\
&=&  N \tfrac{(\eta/2)V^2 \cos(N\Phi_0)^2}{(1+V \sin(N\Phi_0))(1-\eta/2(1+V \sin(N \Phi_0)))}.  \label{Eq:sensi}
\end{eqnarray}
A phase sensitivity of  $\mathcal{S}>1$ then beats the SQL, and the  Heisenberg limit is reached when $\mathcal{S}=\sqrt{N}$.

As eq.(\ref{Eq:sensi}) shows, the phase sensitivity depends on the bias phase $\Phi_0$ in a rather non-trivial manner. Moreover, the dependences of the phase sensitivity on the efficiency $\eta$ and the visibility $V$ are quite different from each other. Fig. 1 illustrates these different dependences for $N=4$. Fig. 1(a) shows the interference fringe $p(\phi)$, as given by eq.(\ref{Eq:prob}).
Fig. 1(b) shows the phase sensitivity $\mathcal{S}$ for various visibilities $V$ at $\eta=1$. As might be expected, the maximal values of $S=\sqrt{N} V$ are obtained where the slope $|dp/d\Phi_0|$ of the fringes is maximal. In the case of $V=0.4$, $\mathcal{S}$ is always smaller than $1$, indicating that the SQL cannot be beaten with $V$ lower than a threshold value of $1/\sqrt{N}=0.5$.
Fig. 1(c) shows $\mathcal{S}$ for various efficiencies $\eta$ at $V=1$. In contrast to Fig. 1(b), maximal sensitivity is now obtained at the minima of the interference fringes $p(\Phi_0)$, even though the slope at these points is equal to zero. The key to understanding this curious phenomena is that, since $V=1$, $p$ is exactly zero at these points. Thus the variance of $C_k$ given by eq.(\ref{Eq:deltaCk}) is also zero, and eq. (\ref{Eq:sensi}) defines a finite phase sensitivity depending on the asymptotic ratio of the squared slope and the variance. The maximal sensitivity thus obtained is $S=\sqrt{N \eta}$, indicating that the SQL cannot be beaten with $\eta$ lower than a threshold value of $1/N=0.25$.
Finally, fig. 1(d) shows the phase sensitivity for $\eta = 3/4$ and $V=0.82$ which correspond to the data obtained in the experiment described below. Interestingly, the combined effects of $V<1$ and $\eta<1$ results in zero sensitivity at the minima of $p(\Phi_0)$, but bends the maxima away from the maximal slope of the fringes and towards the positions of the minima. As a result, the phase bias which gives the maximum phase sensitivity is not the same as the phase with the maximal slope of the event probability $p(\Phi_0)$.
\begin{figure}[t]
\scalebox{0.5}[0.5]{
\includegraphics{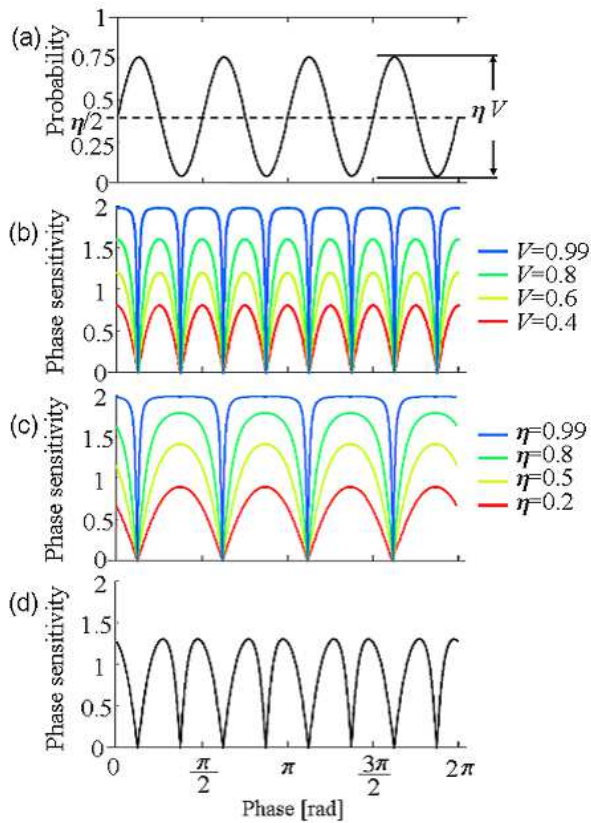}
}
\caption{Phase sensitivity for $N$-photon interference. (a) Detection probability fringe with $N=4$, where $\eta /2$ and $\eta V$ correspond to the average and the amplitude of the fringe, respectively. (b) Phase sensitivities $\mathcal{S}$ for $\eta=1$ and various visibilities $V$. (c) Phase sensitivities $\mathcal{S}$ with $V=1$ and various efficiencies $\eta$. (d) Phase sensitivity $\mathcal{S}$ for $\eta = 3/4$ and $V=0.82$, corresponding to our experimental results.}
\end{figure}

In order to find the phase bias at which optimal phase sensitivity is obtained, it is convenient to express the phase error in terms of the Heisenberg limit plus excess noise,
\begin{equation}
\delta \phi^2 = \tfrac{1}{k N^2}\left(1 + \tfrac{(2-\eta(1+V^2))+2(1-\eta)V \sin(N \Phi_0)}{(\eta V^2(1-\sin(N\Phi_0)^2))}\right). \label{Eq:PhVa}
\end{equation}
By minimizing this function of $\sin(N \Phi_0)$, we can find the following relation for the optimal phase bias $\Phi_{\mbox{\small opt}}$,

\begin{equation}
\sin(N \Phi_{\mbox{\small opt}}) = \tfrac{(\eta/2)(1+V^2)-1+\sqrt{(1-V^2)((1-\eta/2)^2-(\eta V/2)^2)}}{(1-\eta)V}.
\end{equation}
The maximum phase sensitivity obtained at this phase bias is
given by
\begin{equation}
\mathcal{S}_{M}^2 =  N (1+\tfrac{(1-\eta)^2}{\eta (1-(\eta/2)(1+V^2)-\sqrt{(1-V^2)((1-\eta/2)^2-(\eta V/2)^2)})})^{-1}. \label{Eq:Max_Sen}
\end{equation}
Eq. (\ref{Eq:Max_Sen}) defines the phase sensitivity of $N$-photon interference in terms of the experimental parameters $V$ and $\eta$. It thus provides the basis for evaluating the actual phase sensitivities achieved in a specific experiment.
A contour plot of eq. (\ref{Eq:Max_Sen}) is shown in Fig. 2. The values on the right hand side give the sensitivities corresponding to each contour.
\begin{figure}[t]
\scalebox{0.5}[0.5]{
\includegraphics{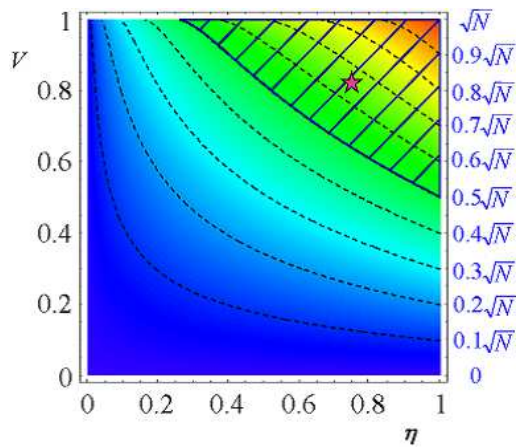}
}
\caption{Contour plot of the maximum phase sensitivity $\mathcal{S}_{M}$. The values on the right end are the sensitivities along the contours. The phase sensitivities in the area covered by diagonal lines beat the SQL for $N=4$. The star indicates the performance of our 4-photon interferometer for the data in Fig. 3.}
\end{figure}
Note that the sensitivities increase with $\sqrt{N}$, indicating that the SQL can be beaten at lower values of $V$ and $\eta$ as the photon number $N$ increases. The area shaded with diagonal lines indicates the region where the sensitivity $\mathcal{S}_{M}$ is greater than one for $N=4$, which is the condition for beating the SQL with a four photon interferometer. Note that our theory does not consider the quantum efficiencies of the single photon detectors or the optical losses. The result therefore represents the phase sensitivity that could be achieved by the N-photon state in the interferometer if no further photons are lost in the detection process.

%
\textit{Experiment}.---We now apply the evaluation of phase sensitivity derived above to an improved version of the four photon interference experiment
reported in \cite{na-sci-316-726}, and proposed in \cite{st-pra-65-033820}. A product state $|22\rangle_{ab}$ of two two-photon Fock states is generated by parametric down-conversion and injected into a Mach-Zehnder (MZ) interferometer (Fig.3 (a)). The state after the first beam splitter of the interferometer is
\begin{equation}
|\psi_{\mbox{path}} \rangle =
\sqrt{\tfrac{3}{8}}|40\rangle_{cd} +\sqrt{\tfrac{1}{4}}|22\rangle_{cd} + \sqrt{\tfrac{3}{8}}|04\rangle_{cd}, \label{Eq:22MZ}
\end{equation}
where $c$ and $d$ are the two paths inside the interferometer. A phase shift of $\phi$ is then applied to mode $d$. The state after the second beam splitter of the interferometer is
\begin{eqnarray}
|\psi_{\mbox{out}} \rangle &=&
\tfrac{\sqrt{6}}{16}(1-2e^{i 2 \phi}+e^{i 4 \phi})(|40\rangle_{ef}+|04\rangle_{ef})
\nonumber \\ && +
\tfrac{1}{8}(3+2e^{i 2 \phi}+3e^{i 4 \phi})|22\rangle_{ef}
\nonumber \\ && +
\tfrac{\sqrt{6}}{8}(1-e^{i 4 \phi}) (|31\rangle_{ef}+|13\rangle_{ef}),
\end{eqnarray}

\noindent where $e$ and $f$ are the output modes of the MZ interferometer.
Significantly, the amplitudes of the $|31\rangle_{ef}$ and $|13\rangle_{ef}$ components do not include the phase oscillation of $\exp(i 2 \phi)$ associated with the $|22\rangle_{cd}$ component inside the interferometer. It is therefore possible to observe pure four photon interference fringes in the detection probabilities $P_{3ef}$ of $|31\rangle_{ef}$ and $P_{3fe}$ of $|13\rangle_{ef}$.
In our previous experiment \cite{na-sci-316-726}, we counted only the detection of three photons in mode $e$ and 1 in mode $f$, for an efficiency of $\eta=3/8$. However, we can also observe a four photon fringe by detecting one photons in mode $e$ and 3 in mode $f$. By using both detection events, we can improve $\eta$ to $3/4$. Ideally, the total detection probability then reads   $P_{3ef+3fe}=\tfrac{3}{8}(1-\cos4\phi)$.

A frequency doubled 780 nm fs pulsed laser (repetition interval 13 ns, power 200mW) pumps a type-I phase-matched Beta Barium Borate (BBO) crystal (2mm thickness) to generate the state $|22\rangle_{ab}$ via spontaneous parametric down-conversion. The down-converted photons pass through interference filters with 4 nm bandwidth and are then guided via polarization maintaining fibres (PMFs) to a displaced-Sagnac interferometer, which is essentially equivalent to a MZ interferometer \cite{na-sci-316-726}. A variable phase shift in mode $d$ is realized by changing the angle of a phase plate (PP) in the interferometer. Photons are collected in single mode fibres (SMFs) at the output modes and detected using three cascade single photon counting modules (SPCM, detection efficiency 60 \% at 780 nm) in modes $e$ and $f$ (a total of six detectors). To test the performance of the four-photon interferometer, we used a relatively low efficiency source and modest efficiency detectors which means that many more photons pass through the interferometer than lead to a four-photon detection event. For applications (such as biological sensing) where photon flux is important, high efficiency number resolving visible light photon counters \cite{ta-apl-99-1063, jun-apl-74-902} would dramatically improve detection efficiency. 
Using a home-made  coincidence counter, we counted either of the following four fold coincidence events: (i) detection signals from one of the three counters in mode $e$ and all of the three counters in mode $f$, and (ii) from one of the three counters in mode $f$ and all of the three counters in mode $e$.

\begin{figure}[t]
\scalebox{0.5}[0.5]{
\includegraphics{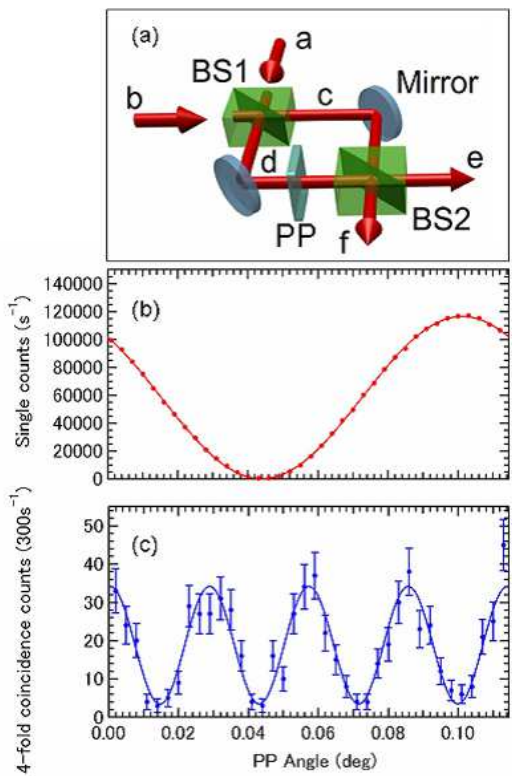}
}
\caption{Experimental $\lambda/4$ fringe with $\eta=3/4$. (a) A schematic of a Mach-Zehnder interferometer consisting of two 50:50 beamspltters (BS1 and BS2). (b) single photon counts as a function of phase plate (PP) angle for a single photon input. (c) four-fold coincidence counts of three photons in mode $e$ (or $f$) and one photon in mode $f$ (or $e$).}
\end{figure}
Figure 3 (b) shows a single-photon interference fringe with $V=99.2 \pm 0.3 \%$, obtained by inputting single photons in mode $a$ and detecting the rate of single photons in mode $e$. The result of the four photon interference ($P_{3ef+3fe}$) is shown in Fig. 3 (c). As expected, the fringe period is $1/4$ that of Fig. 3 (b), demonstrating phase super-resolution. The visibility $V$ of the fitted curve is $82 \pm 6\%$. Using Eq. \ref{Eq:Max_Sen}, we can now determine the phase sensitivity for the experimental parameters $\eta=3/4$ and $V=0.82$. The maximum phase sensitivity achieved by the four photon interferometry is then found to be $\mathcal{S}_{M} = 1.30$,
\textit{i.e.} the phase sensitivity of our interferometer is 1.3 times greater than the SQL. Note that $S_M=1.30$ is the sensitivity value that could be reached if the experiment is performed with unit quantum efficiency detectors and without losses. This experimental result is indicated by the star mark in Fig. 2, which illustrates the relation between the experimental parameters and the conditions for beating the SQL. Note that $\mathcal{S}_{M} = 1.30$ is achievable not at the maximum slope of $p$ but at the points shifted to the valleys of $p$ as shown in Fig.1 (d). 

%
\textit{Summary}.---We have derived the phase sensitivity when the detection probability of an output event is given by an $N$-photon interference fringe. We find that the phase sensitivity shows quite different dependences on the efficiency $\eta$ and the visibility $V$. As a result, the phase bias that gives maximum phase sensitivity is in general not the same as the phase with the maximum slope of the detection probability. We have determined the optimal phase bias and the corresponding maximum phase sensitivity as a function of efficiency $\eta$ and visibility $V$ obtained for a specific experiment. With this result, we can determine the quantitative enhancement of the phase sensitivity achieved in any $N$-photon interference experiment. In order to illustrate the theoretical results, we have applied this evaluation method to an improved four photon interference experiment with an efficiency of $\eta=3/4$ and obtained a maximum phase sensitivity of $1.3$ above the SQL given the experimentally observed visibility of $0.82$. Finally, it may be possible to extend the analysis presented here to include schemes which allow multiple passes of single photons \cite{Hi-nat} and those using trapped ions \cite{bo-pra-54-4649, le-nat-438-639}.

This work was supported by the Japan Science and Technology Agency (JST), Ministry of Internal affairs and Communication (MIC), Japan Society for the Promotion of Science (JSPS), 21st century COE program, Special Coordination Funds for Promoting Science and Technology, UK EPSRC QIP IRC, and the Daiwa Anglo-Japanese Foundation.

\end{document}